\documentstyle[12pt,epsf]{article}
\begin{document}
\title{
{\bf On the Neuberger overlap operator}
}

\author{Artan Bori\c{c}i \\
        {\normalsize\it Paul Scherrer Institute}\\
        {\normalsize\it CH-5232 Villigen PSI}\\
        {\normalsize\it Artan.Borici@psi.ch}\\
}

\date{}
\maketitle

\begin{abstract}
We compute Neuberger's overlap operator by the Lanczos algorithm
applied to the Wilson-Dirac operator.
Locality of the operator for quenched QCD data and its eigenvalue
spectrum in an instanton background are studied.
\end{abstract}


\bigskip
\bigskip

\bigskip
{\bf 1.}
Although brute force calculations of the quenched lattice QCD
with Wilson fermions have been able to approach the
chiral limit
\cite{CPPACS}, there are increased efforts to make the
chiral symmetry exact on the lattice
\cite{Narayanan_Neuberger,Hasenfratz_Laliena_Niedermayer&Luescher1}.
\footnote{For a recent review on the topic see \cite{Niedermayer}.}

There are different starting points to formulate
lattice actions with exact lattice chiral symmetry, but
all of them seem to obey the Ginsparg-Wilson condition
\cite{Ginsparg_Wilson}:
\begin{equation}
\gamma_5 D^{-1} + D^{-1} \gamma_5 = a \gamma_5 {\alpha}^{-1},
\end{equation}
where $a$ is the lattice spacing,
$D$ is the lattice Dirac operator and ${\alpha}^{-1}$ is a local
operator and trivial in the Dirac space.

A candidate is the overlap operator of Neuberger \cite{Neuberger1}:
\begin{equation}
D = 1 - \gamma_5 \mbox{sign}(H), ~~~~H = \gamma_5(\sigma - aD_{W})
\end{equation}
where $\sigma$ is a shift parameter in the range $(0,2)$,
which we have fixed at one and $D_{W}$ is the Wilson-Dirac operator,
\begin{equation}
D_{W} = \frac{1}{2} \sum_{\mu}
[\gamma_{\mu} (\partial_{\mu}^{*} + \partial_{\mu})
            - a\partial_{\mu}^{*}\partial_{\mu}]
\end{equation}
and $\partial_{\mu}$ and $\partial_{\mu}^{*}$ are the nearest-neighbor
forward and backward difference operators.

The locality has been shown for smooth background fields
and no violation has been observed in quenched samples simulated
at moderate couplings
\cite{Hernandez_Jansen_Luescher}.

\bigskip
{\bf 2.} So far, all the methods devised to compute the overlap operator
by usual iterative solvers
have been based on (rational) polynomial approximations of the inverse
square root or the $sign$ function
\cite{Neuberger2,Edwards_Heller_Narayanan}.
(Mathematical foundations of these methods are reviewed in \cite{Higham}.)
But they may
exceed the storage limits in some machines. This is not the case with
Legendre \cite{Bunk} and Chebyshev \cite{Hernandez_Jansen_Luescher}
polynomials, which on the other hand are not optimal \cite{coment}.
\footnote{After submission of this paper, the rational polynomial
approximation method \cite{Neuberger2} was improved with respect to
memory limits \cite{Neuberger3}
by running twice the Conjugate Gradient (CG) iteration,
as it is the case here (see below) for the Lanczos algorithm.}

In the present work we propose a new method, which
uses the outcome of the Lanczos algorithm on $H$.
The Lanczos iteration is known to approximate the spectrum of
the underlying matrix in an optimal way and, in particular,
it requires a constant memory \cite{Golub_VanLoan}.

Let $Q_n = [q_1,\ldots,q_n]$ be the set of orthonormal vectors,
such that
\begin{equation}\label{HQ_QT}
H Q_n = Q_n T_n, ~~~~q_1 = \rho_1 b, ~~~~\rho_1 = 1/||b||_2
\end{equation}
where $T_n$ is a tridiagonal and symmetric matrix.
Here $b$ stands for an arbitrary vector.

By writing down the above decomposition in terms of the vectors
$q_i, i=1,\ldots,n$ and the matrix elements of $T_n$, we arrive at
a three term recurrence that allows to compute these vectors
in increasing order, starting from the vector $q_1$. This is called
the Lanczos algorithm, which constructs a basis for the
so called Krylov subspace: $span(b,Hb,\ldots,H^{n-1}b)$ \cite{Golub_VanLoan}.

In the last equation, it has been assumed that after $n$ steps of
the Lanczos algorithm, the Krylov subspace remains invariant.
The task is the computation of $x = (H^2)^{-1/2} b$. Our method
is based on the following observations:
Let $(H^2)^{-1/2} = f(H^2)$ be a matrix-valued function, for example
Robert's integral formula \cite{Higham}:
\begin{equation}
f(H^2) = \frac{2}{\pi} \int_0^{\infty} dt (t^2 + H^2)^{-1}
\end{equation}
Then, clearly:
\begin{equation}\label{fQ_Qf}
f(H^2) Q_n = Q_n f(T_n^2)
\end{equation}
Since, on the other hand,
\begin{equation}
b = Q_n e_1^{(n)} / \rho_1,
\end{equation}
where $e_1^{(n)}$ denotes the unit vector with $n$ elements
in the direction $1$, we get:
\begin{equation}\label{result}
x = f(H^2) b = Q_n f(T_n^2) e_1^{(n)} / \rho_1
\end{equation}
There are some remarks to be made here:

a) By applying the Lanczos iteration on $H$, the problem
of computing $(H^2)^{-1/2}$ reduces to the problem of computing
$(T_n^2)^{-1/2}$ which is typically a much smaller problem than
the original one. It can be solved for example by using the full
decomposition of $T_n$ in its eigenvalues and eigenvectors; in fact
this is the method we have employed too, for its compactness and
the small overhead for moderate $n$.

b) In the floating point arithmetic, there is a danger that
once the Lanczos polynomial (algorithm) has approximated well some part
of the spectrum,
the iteration reproduces vectors which are rich
in that direction \cite{Simon}.
As a consequence, the orthogonality of the Lanczos vectors is spoiled
with an immediate impact on the history of the iteration.

c) In general, there is no guarantee
that the algorithm will converge at smaller $n$,
unless $n = \mbox{rank}(H)$ in exact arithmetic \cite{Golub_VanLoan}.
Therefore, for a given $n$ the equations (\ref{HQ_QT}) and
(\ref{result}) hold approximately.

Therefore, in practical implementations one should be satisfied
with a stopping criterium such as:
\begin{equation}\label{residual_error}
\mbox{error}(n) = |(\rho_1 ||H x_n||_2)^2 - 1|^{1/2}
\end{equation}
is made small enough.

It is worth writing down the error in terms of the Lanczos matrix;
straightforward algebra gives:
\begin{equation}\label{error}
\mbox{error}(n) = |\beta_n z_n^{(n)}|,
~~~~z_n = (T_n^2)^{-1/2} e_1^{(n)}
\end{equation}
where $\beta_n$ is the element $(n+1,n)$ of the matrix $T_{n+1}$
and $z_n^{(n)}$ is the last component of the vector $z_n$.

Since $H$ and $(H^2)^{1/2}$ are equally conditioned in 2-norm,
we expect, that once the system $H x = b$ is solved, the system
$(H^2)^{1/2} x = b$ is also solved. In this context, it is desirable
to compare the error (\ref{error}) with the residual error
of the original system, $r_n$. As before, in terms of the
Lanczos matrix, it is given by:
\begin{equation}\label{residual}
||\rho_1 r_n||_2 = |\beta_n y_n^{(n)}|,
~~~y_n = T_n^{-1} e_1^{(n)}
\end{equation}
As long as the orthogonality between Lanczos vectors is sufficiently
maintained, equations (\ref{error}-\ref{residual}) should hold
to a good accuracy.

To implement the result (\ref{result}),
we first construct the Lanczos matrix
and then compute $z_n$. By repeating the iteration, we compute
Lanczos vectors and obtain the result. We saved the scalar products,
though it was not necessary.
If we call $1/\rho_i, i = 1, \ldots$ the norm of the residual error
of the system $H x = b$, it is easy to show that
\begin{equation}
\rho_{i+1} \beta_i + \rho_i \alpha_i + \rho_{i-1} \beta_{i-1} = 0
\end{equation}
Therefore, we have the
following {\em algorithm for solving the system $(H^2)^{1/2} x = b$:}
\begin{equation}
\begin{array}{l}
\beta_0 = 0, ~\rho_1 = 1 / ||b||_2, ~q_0 = o, ~q_1 = \rho_1 b \\
for ~i = 1, \ldots \\
~~~~v = H q_i \\
~~~~\alpha_i = q_i^{\dag} v \\
~~~~v := v - q_i \alpha_i - q_{i-1} \beta_{i-1} \\
~~~~\beta_i = ||v||_2 \\
~~~~q_{i+1} = v / \beta_i \\
~~~~\rho_{i+1} = - \frac{\rho_i \alpha_i + \rho_{i-1} \beta_{i-1}}{\beta_i} \\
~~~~if \frac{1}{|\rho_{i+1}|} < tol, ~n = i, ~end ~for \\
\\
Set ~(T_n)_{i,i} = \alpha_i, ~(T_n)_{i+1,i} = (T_n)_{i,i+1} = \beta_i,
otherwise ~(T_n)_{i,j} = 0 \\
z_n = (T_n^2)^{-1/2} = U_n (\Lambda_n^2)^{-1/2} U_n^T e_1^{(n)} \\
\\
q_0 = o, ~q_1 = \rho_1 b, ~x_0 = o \\
for ~i = 1, \ldots, n \\
~~~~x_i = x_{i-1} + q_i z_n^{(i)} / \rho_1 \\
~~~~v = H q_i \\
~~~~v := v - q_i \alpha_i - q_{i-1} \beta_{i-1} \\
~~~~q_{i+1} = v / \beta_i \\
\end{array}
\end{equation}
where by $o$ we denote a vector with zero entries
and $U_n, \Lambda_n$ the matrices of the egienvectors and eigenvalues
of $T_n$. Note that
there are only four large vectors necessary to store: $q_{i-1},q_i,v,x_i$.

Obviously, the memory doesn't grow with $n$.
This is not the case for the shifted CG iterations
(\cite{Neuberger2,Edwards_Heller_Narayanan}) needed to compute
the (rational) polynomial approximation of $(H^2)^{-1/2}$.
Since there is a one to one connection between CG and Lanczos,
such approximations on the original matrix
are transfered to the corresponding Lanczos matrix \cite{Golub_VanLoan}.
This should be contrasted with the exact computation of
$(T_n^2)^{-1/2}$.

To test the above analysis, we have performed simulations of SU(3) gauge
theory at $\beta = 6.0$ on a $8^316$ lattice and picked up an equilibrated
configuration.

In Fig.1
we show the residual error $|(\rho_1 ||H x_n||_2)^2 - 1|^{1/2}$
computed directly from $x_n$ and compare with the same quantity
given in terms of the Lanczos matrix, i.e. $|\beta_n z_n^{(n)}|$.
It fluctuates between two branches: the upper one corresponds to odd
$n$, the number of matrix-vector multiplications, and the lower one
to even values of $n$. For large $n$, the computed and estimated errors
deviate from each other, which may indicate accumulation of roundoff
errors in the computed residual error. Note that we have employed
64 bit precision.

For comparison, we show in Fig.2 the residual error $||\rho_1 r_n||_2$
of the system $H x = b$ as computed directly and from $|\beta_n y_n^{(n)}|$.
Again, we have two branches as explained above, but here there is no
distinction between the computed and estimated errors.
The appearance of two branches is not surprising since we are dealing
with a non-definite matrix $H$.

In Fig.3 we compare the computed residual errors of the systems
$H x = b$ and $(H^2)^{1/2} x = b$ (upper branches).
They are the same most of the time, unless $n$ becomes large and
deviations become clearer. This behavior shows that both systems
are solved at the same time, which should serve us as a guide, because
the computation of $x_i$ at each step $i$ is very demanding.

We have compared the efficiency of the above method with that of
the rational polynomial approximation \cite{Neuberger2}.
First, we solved the system $H x = b$
by both Lanczos and CG with an $10^{-5}$ accuracy for the
residual error. We needed the same number
of multiplications with $H^2$, 226. To compute the inverse square root
with the same accuracy ($10^{-5}$),
we applied then the above method and the rational polynomial method
with $N = 30$, the number of the terms in the approximating
sum (\cite{Neuberger2}).
(Smaller $N$ will give a lower accuracy: in Fig. 4 we display
the norm of the residual error (\ref{residual_error}) as a function
of $N$.) Therefore, if we stored the Lanczos vectors as in the
rational polynomial method, both methods need about the same
amount of work. To avoid memory restrictions, we increase the work
by a factor of two.
\footnote{As it was stated above,
after submission of this paper, the rational polynomial
approximation method \cite{Neuberger2} was changed in the same
fashion, by increasing the work by a facor of two \cite{Neuberger3}.}

\bigskip
{\bf 3.}  An immediate application of the above method is to check
the locality of Neuberger's overlap operator $D$.

We used 100 equilibrated
SU(3) configurations at $\beta = 5.7,6.0$ on a $8^316$ lattice.
For each configuration we computed the absolute value of
$D$ elements in the first column with space, spin and color
indices fixed at one, a selection that suffices to look for
violations of the locality.

The average over 100 configurations is plotted in Fig. 5.
At both values of $\beta$, there was no
single configuration to show an exceptional
behavior: the maximum deviation is slightly above the mean value.
The values of $D$ decrease rapidly with the time slices. For large $t/a$
but away from the center,
\footnote{Throughout the paper
we have used periodic boundary conditions in all directions.}
they fall off exponentially.

Recently, it has been shown that for $\beta = 6.0,6.2,6.4$
the occurrence of configurations with exceptionally small eigenvalues
of $H^2$ becomes vanishingly small for $12^4$ and $16^4$ large
lattices, whereas for sufficiently smooth gauge fileds the locality
is guaranteed \cite{Hernandez_Jansen_Luescher}.

In fact at $\beta = 5.7$ the complete
spectrum of $1 - D_{W}$ is in the left half
of the complex plane.
\footnote{I am grateful to Urs M. Heller for pointing out that
normal spectroscopy with Wilson fermions can be done
at $\beta = 5.7$ and $\kappa = 0.1675$ \cite{GF11},
the latter being greater than our corresponding $\kappa = 1/6$.}
We computed partial spectra of $\gamma_5 H$ by the implicitely restarted
Arnoldi iteration \cite{Sorensen} and found that indeed at this coupling
there was no eigenvalue in the right half of the complex plane for all
our 100 configurations. Therefore,
there exist a $\beta \in (5.7,6.0)$ at which the operator $1 - D_{W}$
becomes singular.
In this case, there is an unbounded number of near zero modes of $H$
and therefore $D$ is no longer local.

\bigskip
{\bf 4.} As another application we consider the lattice index theorem for
an SU(2) instanton background on a small $4^4$ lattice.
An instanton on the lattice
can be prepared in various ways. We follow \cite{Laursen_Smit_Vink}
and prepare an instanton with size $\rho$
in the center of the lattice in the singular gauge.
The index of $D$ is given by:
\begin{equation}
\mbox{index}(D) = \frac{1}{2} \mbox{Tr}[\mbox{sign}(H)]
\end{equation}
Because of the O(a) lattice errors, we expect the instanton being observed
for $\rho \geq a$.
We computed the smallest eigenvalues of $D$
by an (implicitly restarted) Arnoldi iteration with the non-converged
Ritz values used as explicit shifts \cite{Sorensen}. We fixed the
number of Arnoldi steps at 32 and have stopped the iteration when the
next starting vector norm is smaller than $10^{-6}$. We have checked
the stability
of the computed eigenvalues by increasing the cutoff beyond
the number of the converged eigenvalues. The stability is observed unless
the cutoff becomes too large, which means that a larger Arnoldi matrix
should be employed.

We note that it is crucial for the eigenvalue computation to have a
proper accuracy in the computation of $D$, which in our case has
been set at a residual error norm (\ref{residual_error}) less than $10^{-10}$.

We have computed eigenvalues for $\rho/a \in[0.5,1.5]$ with steps of $0.1$.
For brevity, we show in Table 1 eigenvalues of a smaller set of $\rho$.
For $\rho \leq a$ there are two zero modes,
whereas for $\rho > a$ there is a single zero mode present. As numerical
accuracy is an issue here, we have perturbed the instanton background by
applying a small fluctuating gauge field. The picture doesn't change, but
the zero modes for $\rho \leq a$ become nearly
zero modes with opposite chiralities.

We have also computed the eigenvalues exactly by standard QR algorithms.
The instanton mode appears single in both approaches. While the values of
the other smallest eigenvalues are reproduced exactly
by the implicitly restarted
Arnoldi algorithm, their multiplicity cannot be handled. Since $D$ is
normal, the Arnoldi matrix is normal and tridiagonal and therefore
irreducible, giving no information on the multiplicity. The latter is essential
when $D$ has more exact zero modes and one must rely on
block variants of the same algorithm.

Even on such a small lattice, the computation of the zero modes is
not the fastest method to compute the topological charge. Tracing the
crossings of the smallest eigenvalues of $H$ is more practical.

Nonetheless we note that an estimation of the topological charge can
be made during the computation of $D$ as described 
in this work. Having computed
the Lanczos matrix of $H$ one can estimate sign($H$), which for our
SU(2) instanton gives excellent agreement for most of starting vectors $b$.
In general, a separate study is needed to conclude on this method.

\bigskip
{\bf 5.} To conclude:
we have computed with a new method the overlap operator based on the Lanczos
algorithm applied on the Wilson-Dirac operator. Compared to the other
methods \cite{Neuberger2,Edwards_Heller_Narayanan}, its main advantage
is of being free from memory restrictions. Additionally, there is no
approximation made in the computation of the inverse square root of the
Lanczos matrix.

The locality of the overlap operator has been tested. We recommend to
check it always before any other computation.

The computation of $D$ turns to be more difficult than $D_{W}$.
However, the so-called classically perfect actions
\cite{Niedermayer}
may help substantially to work on moderate lattices.
Further studies are needed for the dynamical implementation of $D$.

We are grateful to stimulating discussions with Ferenc Niedermayer on the
topics covered by this work and to
Roland Rosenfelder and Philippe de Forcrand
for making critical remarks on this paper.

We are grateful to Herbert Neuberger for comments on the method presented
in this work following the posting of its first version.

We thank PSI where this work was done and SCSC Manno
for the allocation of computer time on the NEC SX4.

\pagebreak

\begin{figure}
\epsfxsize=12cm
\epsfxsize=10cm
\vspace{3cm}
\centerline{\epsffile[100 200 500 450]{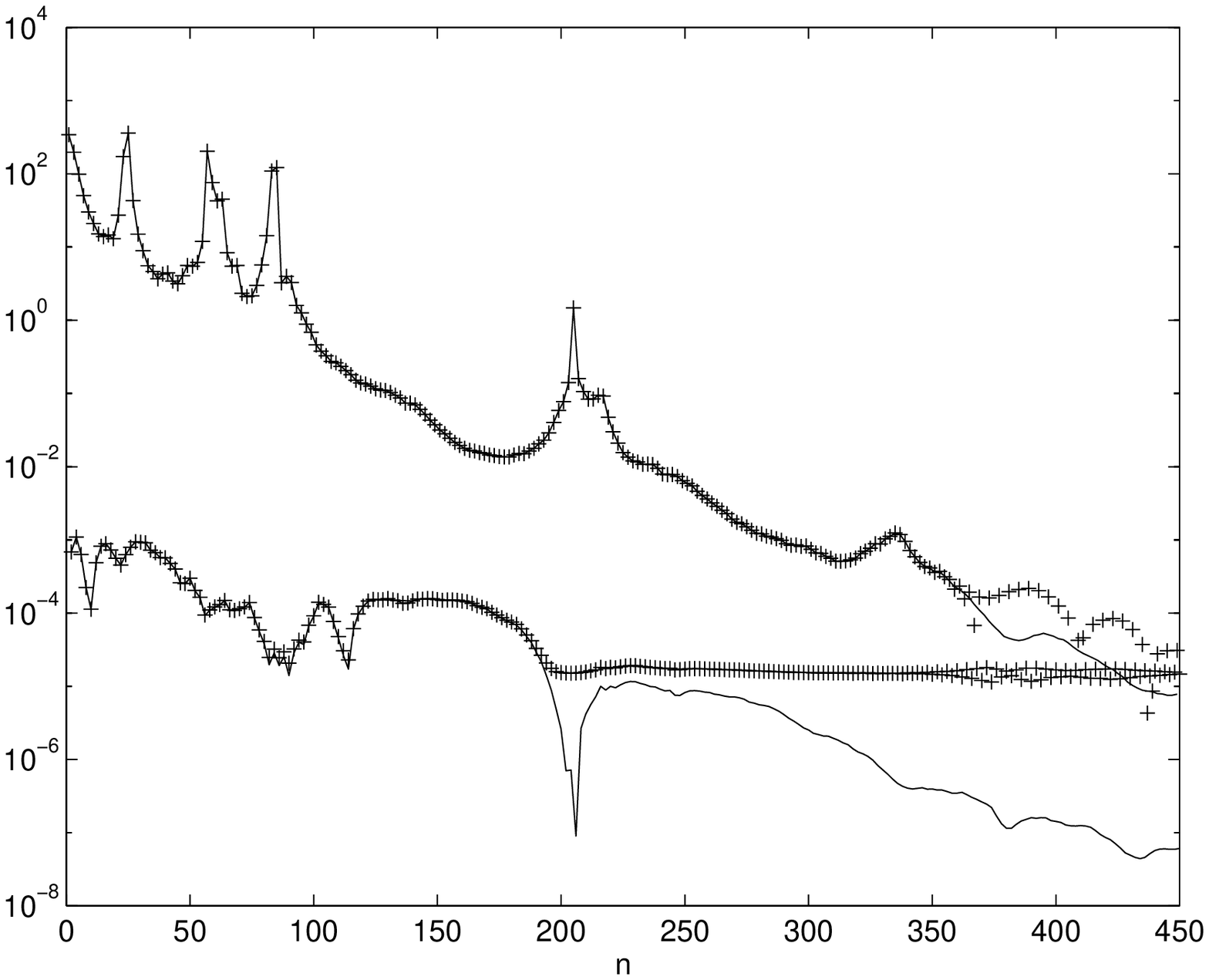}}
\caption{Residual error of the system $(H^2)^{1/2} x = b$
as defined in (\ref{residual_error}): plus symbols that jump
between two branches.
The same quantity defined in terms of the Lanczos matrix is
displayed by the solid line.}
\end{figure}
\begin{figure}
\epsfxsize=12cm
\epsfxsize=10cm
\vspace{3cm}
\centerline{\epsffile[100 200 500 450]{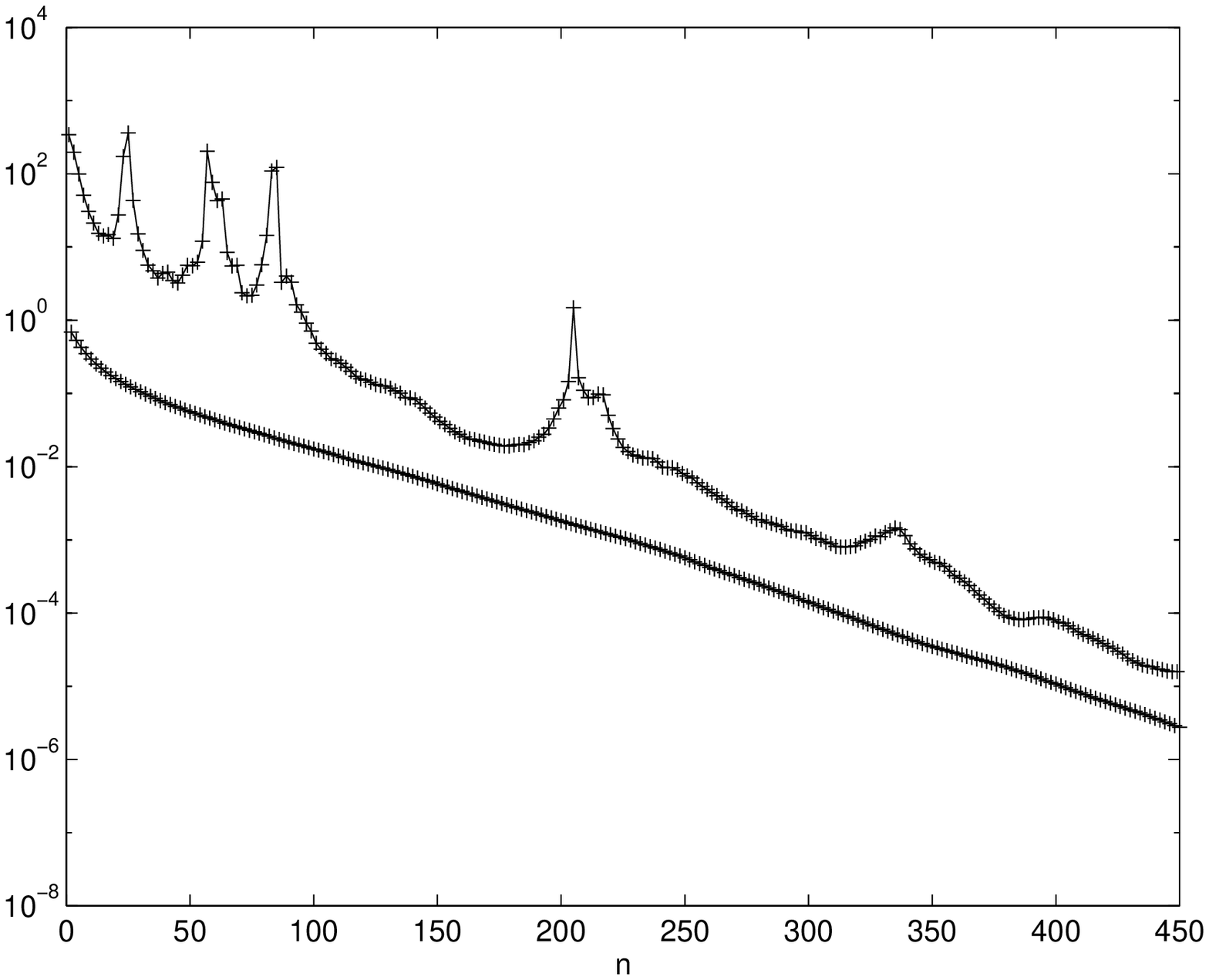}}
\caption{Residual error of the system $H x = b$:
plus symbols that jump
between two branches. 
The same quantity defined in terms of the Lanczos matrix is
displayed by the solid line.}
\end{figure}
\begin{figure}
\epsfxsize=12cm
\epsfxsize=10cm
\vspace{3cm}
\centerline{\epsffile[100 200 500 450]{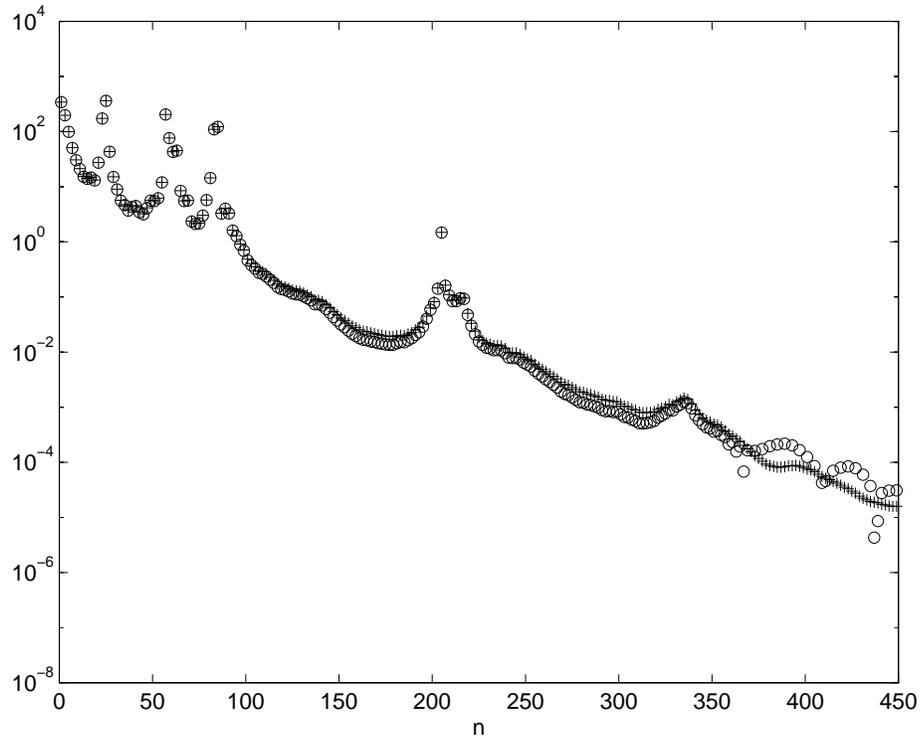}}
\caption{Residual error of the systems $H x = b$ (plus symbols)
and $(H^2)^{1/2} x = b$ (circle symbols) for odd $n$.}
\end{figure}
\begin{figure}
\epsfxsize=12cm
\epsfxsize=10cm
\vspace{3cm}
\centerline{\epsffile[100 200 500 450]{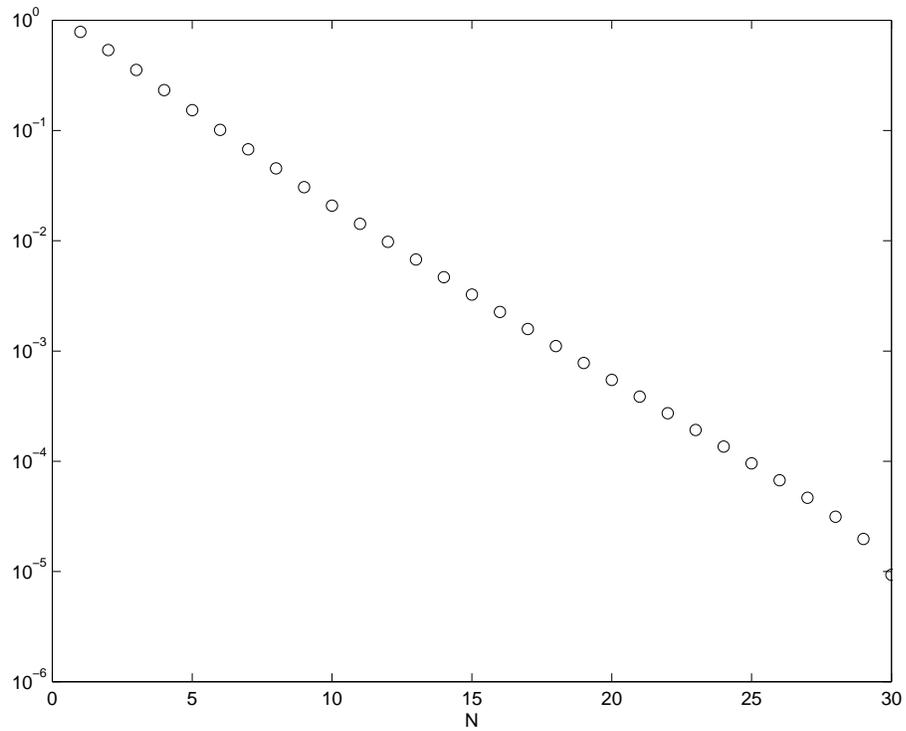}}
\caption{Residual error of the system $(H^2)^{1/2} x = b$
as a function of $N$ in the rational polynomial approximation.}
\end{figure}
\begin{figure}
\epsfxsize=12cm
\epsfxsize=10cm
\vspace{3cm}
\centerline{\epsffile[100 200 500 450]{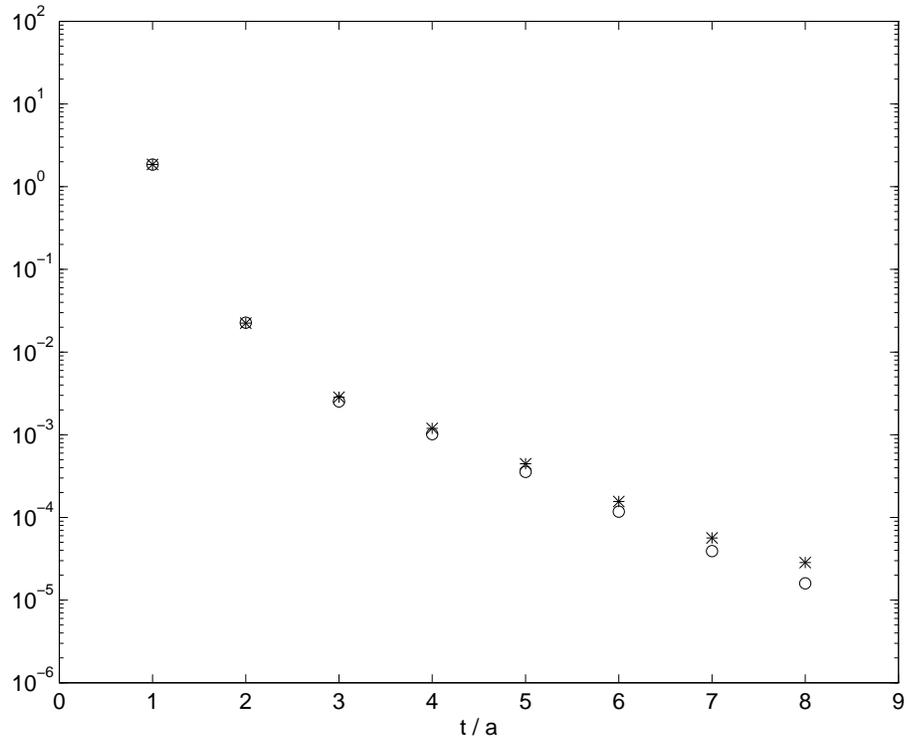}}
\caption{Absolute value of $D$ elements in the first
column with space, spin and color indices fixed at one, as a function
of time indices at $\beta = 6.0$ (circles) and $\beta = 5.7$ (stars).}
\end{figure}
\begin{table}
\begin{center}
\begin{tabular}{|c|c|c|}
\hline\hline
$\rho$ = 0.9a & $\rho$ = 1.0a & $\rho$ = 1.1a \\
\hline
& & 0.277E~-11+i0.660E~-13 \\
 0.124E~-12~-i0.198E~-14 & 0.572E~-11~-i0.694E~-11 & 0.303E+00~-i0.717E+00 \\
 0.955E~-11+i0.167E~-13 & 0.575E~-11+i0.697E~-11 & 0.303E+00+i0.717E+00 \\
 0.947E+00~-i0.998E+00 & 0.940E+00~-i0.998E+00 & 0.303E+00~-i0.717E+00 \\
 0.947E+00+i0.998E+00 & 0.940E+00+i0.998E+00 & 0.303E+00+i0.717E+00 \\
 0.957E+00~-i0.997E+00 & 0.950E+00~-i0.995E+00 & 0.931E+00~-i0.997E+00 \\
 0.957E+00+i0.997E+00 & 0.950E+00+i0.995E+00 & 0.931E+00+i0.997E+00 \\
 0.102E+01~-i0.999E+00 & 0.102E+01~-i0.999E+00 & 0.939E+00~-i0.997E+00 \\
 0.102E+01+i0.999E+00 & 0.102E+01+i0.999E+00 & 0.939E+00+i0.997E+00 \\
 0.104E+01~-i0.998E+00 & 0.104E+01+i0.997E+00 & 0.103E+01~-i0.998E+00 \\
 0.104E+01+i0.998E+00 & 0.104E+01~-i0.997E+00 & 0.103E+01+i0.998E+00 \\
\hline\hline
\end{tabular}
\end{center}
\caption{Smallest eigenvalues of $D$ for three instanton sizes $\rho$}
\end{table}


\begin{thebibliography}{9}

\bibitem{CPPACS}
         R. Burkhalter for the CP-PACS Collaboration,
         {\it Recent Results from the CP-PACS Collaboration},
         UTCCP-P-52, Oct. 1998, and {\tt hep-lat/9810043}.

\bibitem{Narayanan_Neuberger}
         R. Narayanan, H. Neuberger,
         Phys. Lett. B 302 (1993) 62,
         Nucl. Phys. B 443 (1995) 305.

\bibitem{Hasenfratz_Laliena_Niedermayer&Luescher1}
         P. Hasenfratz, V. Laliena and F. Niedermayer,
         Phys. Lett. B 427 (1998) 125,
         M. L\"uscher,
         Phys. Lett. B 428 (1998) 342.

\bibitem{Niedermayer}
         F. Niedermayer,
         {\it Exact chiral symmetry, topological charge and
          related topics},
         {\tt hep-lat/9810026}.

\bibitem{Ginsparg_Wilson}
         P. H. Ginsparg and K. G. Wilson,
         Phys. Rev. D 25 (1982) 2649.

\bibitem{Neuberger1}
         H. Neuberger,
         Phys. Lett. B 417 (1998) 141,
         Phys. Rev. D 57 (1998) 5417.

\bibitem{Hernandez_Jansen_Luescher}
         P. Hern\'andez, K. Jansen and M. L\"uscher,
         CERN-TH/98-250, DESY 98-094, and {\tt hep-lat/9808010}.

\bibitem{Neuberger2}
         H. Neuberger,
         Phys. Rev. Lett. 81 (1998) 4060.

\bibitem{Edwards_Heller_Narayanan}
         R. G. Edwards, U. M. Heller and R. Narayanan,
         {\it A study of practical implementations of the
          Overlap-Dirac operator in four dimensions},
         FSU-SCRI-98-71, and {\tt hep-lat/9807017}.

\bibitem{Higham}
         N. J. Higham,
         Proceedings of
         "Pure and Applied Linear Algebra: The New Generation",
         Pensacola, March 1993.

\bibitem{Bunk}
         B. Bunk,
         Nucl.Phys.Proc.Suppl. B63 (1998) 952.

\bibitem{coment}
         While Legendre polynomials are not optimal
         in any sense here, Chebyshev polynomials are not
         optimal in the sense that their roots do not represent
         the actual distribution of the eigenvalues of $H^2$
         \cite{Golub_VanLoan}.

\bibitem{Neuberger3}
         H. Neuberger,
         {\tt hep-lat/9811019}.

\bibitem{Golub_VanLoan}
         G. H. Golub and C. F. Van Loan,
         {\it Matrix Computations}, The Johns Hopkins University
         Press, Baltimore, 1989.

\bibitem{Simon}
         H. D. Simon,
         Linear Algebra and its Applications 61:101-131(1984).

\bibitem{GF11}
         F. Butler, {\it et al.},
         Nucl. Phys. B 430 (1994) 179.

\bibitem{Sorensen}
         D. C. Sorensen,
         SIAM J. Matrix Anal. Appl. 13 (1992), 357-385.

\bibitem{Laursen_Smit_Vink}
         M. L. Laursen, J. Smit and J. C. Vink,
         Nucl. Phys. B 343 (1990) 522.
       
\end{thebibliography}
\end{document}